\documentclass[aps]{revtex4}
\usepackage{amssymb,epsf}
\usepackage[normalem]{ulem}

\begin{document}

\title{Magnetic solutions in Einstein-massive gravity with linear and
nonlinear fields}
\author{Seyed Hossein Hendi$^{1,2}$\footnote{
email address: hendi@shirazu.ac.ir}, Behzad Eslam Panah$^{1,2}$\footnote{%
email address: behzad.eslampanah@gmail.com} Shahram Panahiyan$^{1,3,4}$%
\footnote{%
email address: shahram.panahiyan@uni-jena.de}, and Mehrab Momennia$^{1}$%
\footnote{%
email address: m.momennia@shirazu.ac.ir},}
\affiliation{$^{1}$ Physics Department and Biruni Observatory, College of Sciences,
Shiraz University, Shiraz 71454, Iran\\
$^2$ Research Institute for Astronomy and Astrophysics of Maragha (RIAAM),
Maragha, Iran\\
$^3$Helmholtz-Institut Jena, Fr\"{o}belstieg 3, D-07743 Jena, Germany\\
$^4$ Physics Department, Shahid Beheshti University, Tehran 19839, Iran}

\begin{abstract}
The solutions of $U(1)$ gauge-gravity coupling is one of the interesting
models for analyzing the semi-classical nature of spacetime. In this regard,
different well-known singular and nonsingular solutions have been taken into
account. The paper at hand investigates the geometrical properties of the
magnetic solutions by considering Maxwell and power Maxwell invariant (PMI)
nonlinear electromagnetic fields in the context of massive gravity. These
solutions are free of curvature singularity, but have a conic one which
leads to presence of deficit/surplus angle. The emphasize is on
modifications that these generalizations impose on deficit angle which
determines the total geometrical structure of the solutions, hence,
physical/gravitational properties. It will be shown that depending on the
background spacetime (being anti de Sitter (AdS) or de Sitter (dS)), these
generalizations present different effects and modify the total structure of the
solutions differently.
\end{abstract}

\maketitle

\section{Introduction}

Existence of topological defects have been reported in various aspects of
the physics and their important roles in physical properties of the systems
have been highlighted. From gravitational/cosmological point of view, the
effects and importance of the topological defects could be related to their
role as a possible dark matter source \cite{dark1,dark2}, their role in
large scale structure of the universe \cite{Kibble1,Kibble2,Kibble6}
anisotropy in the Cosmic Microwave Background (CMB) \cite{AC1,AC4} and their
lensing properties \cite{lens} (which are due to existence of deficit
angle). Essentially, the topological defects in cosmology are produced due
to symmetries that are broken in phase transition that has taken place in
the early universe \cite{Kibble1,Kibble2,Kibble6}. Depending on the number
and type of the symmetries that are broken, these topological defects are
categorized into domain walls (a discrete symmetry is broken and it divides
the universe into blocks), cosmic strings (axial or cylindrical symmetry is
broken and have applications in regard to grand unified particle physics
models/electroweak scale), monopoles (a spherical symmetry is broken) and
textures (several symmetries are broken). Since these topological defects
may be formed during the early universe, they may also carry valuable
information of this era which highlights yet another importance of studying
them. The topological defects are located at the boundaries of regions which
have chosen different minima during the early universe phase transition. So
far, these topological defects have inspired large number of publications
which among them one can point out; cosmic strings in the presence of
Maxwell theory \cite{OJCDias1,OJCDias2}, their superconducting property in
the presence of different models of gravity (such as Einstein \cite{Witten1}%
, Brans-Dicke \cite{AASen} and dilaton gravity \cite{CNFerreira1}), the QCD
\cite{QCD1} and quantum \cite{Quantum1} applications of the magnetic
strings, limits on the cosmic string tension using CMB temperature
anisotropy maps \cite{CMB}, gravitational waves produced by cosmic strings
\cite{spectrum} and decaying domain walls \cite{gravwave} and localization
of fields and chiral spinor on domain walls \cite{local} (For further
studies regarding topological defects, we refer the reader to an incomplete
list of Refs. \cite%
{cosmic1,Skenderis1,Skenderis3,Skenderis5,cosmic8,Skenderis9}). Motivated by
these studies and their interesting results, here we investigate a type of
topological defects which are known as magnetic branes (generalization of
magnetic string), in the presence of two generalizations; massive gravity
and nonlinear electromagnetic field which are generalizations in
gravitational and matter field sectors, respectively.

Although the Maxwell electrodynamics (linear electrodynamics) is one of the
most successful theories in the history of physical science, it does not
provide very precise results in some scales. On the other hand, due to the
fact that the most physical systems are nonlinear in the nature, the
generalization of linear electrodynamics to nonlinear ones seems to be
logical. In addition, owing to specific properties of nonlinear
electrodynamics in the gauge/gravity coupling, the relations between the
general relativity (GR) and nonlinear electrodynamics attract significant
attention. Nonlinear electrodynamic theories have some interesting results
and predictions, and therefore, various nonlinear models of electrodynamics
have been introduced by many authors. For typical examples, one may look at
the Born-Infeld theory \cite{BornInfeld}, logarithmic form \cite{Soleng},
exponential Lagrangian \cite{Hendi}, arcsin nonlinear electrodynamics \cite%
{KruglovI,KruglovII} and etc. Black hole and magnetic solutions by
considering these nonlinear electrodynamics have been investigated in Refs
\cite%
{BHI,BHIII,BHIV2,BHV2,BHV3,BHVI,BHVII2,BHIX,BHIX2,BHXI2,BHXI3,BHXIII,BHXIII2,BHXIV,BHXV2,BHXVI,BHXVI2,BHXVI4,BHXVII,BHXVIII,BHXIX2,BHXX2}%
. Also, other aspects of these nonlinear models have been studied in the
context of quantum level \cite%
{nonlinearQuantum1,nonlinearQuantum2,nonlinearQuantum3,nonlinearQuantum4}
and astrophysical area \cite{nonlinearAstro1,nonlinearAstro2,nonlinearAstro3}
as well. The extensive usages of these nonlinear theories provide the
validity of their authenticity.

Taking into account the conformal invariance, one may find that it plays an
important role in the structure of the some interesting models of string
theory. In other words, conformal invariance is a kind of criterion for
obtaining covariant equations of motion for the on-shell classical
background in the low energy effective of string theory \cite%
{ConformalA,ConformalB}. Regarding the equations of motion in the classical
Einstein gravity, one finds the conformal invariance is equivalent to the
existence of a traceless stress-energy tensor. It is evident that the
Maxwell theory enjoys the conformal invariance only in $4-$dimensions. But
in three or higher dimensional spacetime, the conformal symmetry will be
broken. In other words, the stress-energy tensor of Maxwell theory is
traceless only for $4-$dimensions. In order to keep the conformal invariance
symmetry in arbitrary dimensions, one should generalize the Maxwell field to
the so-called power Maxwell invariant (PMI) theory. PMI theory is one of the
interesting branches of nonlinear electrodynamics in which its Lagrangian is
an arbitrary power of Maxwell Lagrangian ($\mathcal{L}_{PMI}(\mathcal{F}%
)\propto \mathcal{F}^{s}$, where $\mathcal{F}$ is the Maxwell invariant)
\cite{PMII,PMIII,PMIIII,PMIIV}. This theory of nonlinear electrodynamics has
more interesting properties with regard to linear electrodynamics (Maxwell
case), and for the case of $s=1$, it reduces to the Maxwell theory. Another
attractive property is related to conformal invariance property. In other
words, when the power of the Maxwell invariant is a quarter of spacetime
dimensions ($s=d/4$, where $d$ is dimensions), this theory enjoys conformal
invariance, and therefore, its energy-momentum tensor will be traceless. In
this case, one may obtain the Reissner-Nordstr\"{o}m like solutions in higher
dimensions \cite{PMII}. The effects of considering PMI source for the
classical black hole solutions in various gravities have been studied in
literature, for example; Lovelock and Lifshitz black holes with PMI field
have been investigated in \cite{PMIIV,HendiGBPMI,Mazharimousavi,Alvarez},
BTZ black holes in the Einstein and $F(R)$ gravities with this nonlinear
electrodynamic model have been studied in Refs. \cite%
{BTZPMII,BTZPMIII,BTZPMIIII}, the effects of PMI for BTZ black hole with a
scalar hair in the Einstein gravity are reported in \cite{Xu}.
Thermodynamics of topological black holes in the Brans-Dicke theory in the
presence PMI field has been studied before \cite{BDPMI}. Geometrical
thermodynamics and the van der Waals like phase transition of black holes in
higher dimensional spacetimes with PMI theory have been evaluated in
Einstein and dilaton gravity \cite{HendiV,Arciniega,HendiEPT,Mo}. Moreover,
holographic superconductors and magnetic branes (string) supported by PMI
source have been investigated in Refs. \cite{Jing,MagPMII,MagPMIIII}.

Although most of physicists believe that we should respect to conformal
invariance symmetry, they believe that Lorentz invariance symmetry should be
broken in high energy regimes. Considering a nonzero mass for the gravitons
may leads to such breaking symmetry. Recent observations of gravitational
waves from a binary black hole merger provided a firmly evidence of Einstein
theory \cite{Abbott}. However, graviton in Einstein gravity is a massless
particle, whereas there are several arguments that state graviton may be a
massive object \cite{Abbott}. Therefore, GR can be generalized to include
massive gravitons. The first attempt for such generalization was done by
Fierz and Pauli \cite{Fierz} by using a linear theory. However, propagators
of this theory do not reduce to those of GR in limit of vanishing graviton
mass, $m=0$ (van Dam, Veltman and Zakharov discontinuity). In order to
remove this substantial problem, Vainshtein introduced a mechanism which
requires the system to be considered in a nonlinear regime \cite{Vainshtein}%
. Nonetheless, we encounter with Boulware-Deser ghost in the generalization
of Fierz and Pauli massive theory to the nonlinear regime \cite{BDghost}. To
solve such problem, another class of massive gravity was proposed by de
Rham, Gabadadze and Tolley (dRGT) \cite{dRGTI,dRGTII}. dRGT massive theory
is free of Boulware-Deser ghost and it can be used in higher dimensions with
admissible validity \cite{HassanI,HassanII}. It is noteworthy that, in order
to obtain exact solutions with massive terms, an additional metric (called
the reference metric) is invariably needed. The reference metric is required
due to the fact that the interaction terms that can be formed from the
metric alone, cannot be used to construct a mass term. In addition, this is
an unphysical metric that does not have a direct influence on the
geometrical nature of spacetime and it just helps us to find exact solutions
and get rid of Boulware-Deser ghost instability. Considering the suitable
reference metric, one finds various interesting publications in the context
of dRGT massive gravity. Relativistic stars and black object solutions in
dRGT massive gravity with interesting results have been investigated in \cite%
{dGRTBHII,dGRTBHIII,dGRTBHVI,dGRTBHVII,dGRTBHIX}. On the other hand, it was
shown that massive gravity can be expressible on an arbitrary reference
metric \cite{HassanII}. Therefore, a modification in the reference metric
could lead to another dRGT like massive theory. In this respect, Vegh has
introduced a new reference metric with broken translational symmetry\
property \cite{Vegh}. In this massive gravity, similar to dRGT theory,
massive terms are built by using this kind of reference metric which has an
additional property. Also, in this theory, graviton may behave like a
lattice and exhibits a Drude peak \cite{Vegh}, and it is stable and free of
ghost \cite{Zhang}. Neutron stars have been studied in this theory and it
was found that the maximum mass of the neutron stars can be more than $%
3M_{\odot }$ ($M_{\odot }$ is mass of the Sun) \cite{HendiBEP}. Black hole
solutions and its thermodynamic properties have been investigated in Refs.
\cite{VeghBHI,VeghBHII,VeghBHIII}. Besides, the generalizations of this
massive theory to include higher derivative gravity \cite{HendiPEJHEP}, and
gravity's rainbow extension \cite{HendiEPPLB} have been studied as well. In
addition, black hole and magnetic solutions with (non)linear electrodynamics
have been explored in the context of massive gravity \cite%
{BTZHEPJHEP,MagBTZMass}.

In this paper, we want to study the magnetic solutions of Einstein-massive
gravity with linear and nonlinear electrodynamics in four and higher
dimensions. This paper is one of interesting papers for considering the
effects of massive gravitons on the horizonless solutions of nonsingular
spacetime. Before proceeding, we provide some brief motivations for
considering arbitrary higher dimensional spacetimes. In the $20$th century,
Kaluza and Klein introduced a new theory of gravity in five dimensions which
unified gravitation and electromagnetism \cite{Kaluza,Klein}. In addition,
development of string and M-theories led to further progresses in higher
dimensional gravity. Another motivation originates from the anti de
Sitter/conformal field theory (AdS/CFT) correspondence which relates the
properties of $d$-dimensional black holes with quantum field theory in ($d-1$%
)-dimensional hypersurface \cite{Aharony}. On the other hand, with respect
to gravitational researches, one often considers the number of spacetime
dimensions as a free parameter of the theory and investigate its effects.
Studying the effects of this parameter on various aspects of each theory,
may lead to new insights (for example we refer the reader to the effects of
various dimensions in PMI theory \cite{PMII,PMIII,PMIIII,PMIIV}).

The outline of the paper is as follow; in Sec. \ref{FE}, we introduce the
massive gravity with PMI theories and related field equations, briefly. In
Sec. \ref{Sol}, we obtain the solutions in Einstein-Maxwell-massive gravity
and show that these solutions are not black holes, but they contain a conic
singularity. Then, we investigate the effects of all parameters on the
deficit angle. In the next section, we extend the Maxwell source to
nonlinear PMI theory and study the properties of the obtained solutions in
this case, extensively. The last section is devoted to some closing remarks.

\section{Basic Field Equations}

\label{FE}

The $d$-dimensional action in Einstein-massive gravity coupled to
electromagnetic field is given by
\begin{equation}
\mathcal{I}_{G}=-\frac{1}{16\pi }\int_{\mathcal{M}}d^{d}x\sqrt{-g}\left[
\mathcal{R}-2\Lambda +\mathcal{L}(\mathcal{F})+m^{2}\sum_{i=1}^{4}c_{i}%
\mathcal{U}_{i}(g,f)\right] ,  \label{Action}
\end{equation}%
where $\mathcal{R}$\ is the scalar curvature, $\Lambda =\pm
(d-1)(d-2)/2l^{2} $ is the negative/positive cosmological constant for
asymptotically AdS/dS solutions, and $\mathcal{L}(\mathcal{F})$ is an
arbitrary Lagrangian of electrodynamics. In addition, $f$ is a fixed
symmetric tensor, $c_{i}$'s are massive coefficients, and $\mathcal{U}_{i}$%
's are symmetric polynomials of the eigenvalues of matrix $\mathcal{K}_{\nu
}^{\mu }=\sqrt{g^{\mu \alpha }f_{\alpha \nu }}$
\begin{eqnarray}
\mathcal{U}_{1} &=&\left[ \mathcal{K}\right] ,\;\;\;\;\;\mathcal{U}_{2}=%
\left[ \mathcal{K}\right] ^{2}-\left[ \mathcal{K}^{2}\right] ,\;\;\;\;\;%
\mathcal{U}_{3}=\left[ \mathcal{K}\right] ^{3}-3\left[ \mathcal{K}\right] %
\left[ \mathcal{K}^{2}\right] +2\left[ \mathcal{K}^{3}\right] ,  \nonumber \\
&&\mathcal{U}_{4}=\left[ \mathcal{K}\right] ^{4}-6\left[ \mathcal{K}^{2}%
\right] \left[ \mathcal{K}\right] ^{2}+8\left[ \mathcal{K}^{3}\right] \left[
\mathcal{K}\right] +3\left[ \mathcal{K}^{2}\right] ^{2}-6\left[ \mathcal{K}%
^{4}\right] .  \nonumber
\end{eqnarray}

Now, we can obtain the field equations by using variational principle.
Varying the action (\ref{Action}) with respect to both metric tensor and
gauge potential, one can obtain the following field equations
\begin{equation}
R_{\mu \nu }-\frac{1}{2}g_{\mu \nu }\left( \mathcal{R}-2\Lambda \right)
+m^{2}\chi _{\mu \nu }=T_{\mu \nu },  \label{Field equation}
\end{equation}%
\begin{equation}
\partial _{\mu }\left( \sqrt{-g}\mathcal{L}_{\mathcal{F}}F^{\mu \nu }\right)
=0,  \label{Maxwell equation}
\end{equation}%
where $\mathcal{L}_{\mathcal{F}}=d\mathcal{L}(\mathcal{F})/d\mathcal{F}$ and
$\mathcal{F}=F_{\mu \nu }F^{\mu \nu }$\ is the Maxwell invariant in which $%
F_{\mu \nu }$\ $=\partial _{\mu }A_{\nu }-\partial _{\nu }A_{\mu }$ is the
Faraday tensor and $A_{\mu }$ is the gauge potential. In addition, $\chi
_{\mu \nu }$ is the massive term with the following form
\begin{eqnarray}
\chi _{\mu \nu } &=&-\frac{c_{1}}{2}\left( \mathcal{U}_{1}g_{\mu \nu }-%
\mathcal{K}_{\mu \nu }\right) -\frac{c_{2}}{2}\left( \mathcal{U}_{2}g_{\mu
\nu }-2\mathcal{U}_{1}\mathcal{K}_{\mu \nu }+2\mathcal{K}_{\mu \nu
}^{2}\right) -\frac{c_{3}}{2}(\mathcal{U}_{3}g_{\mu \nu }-3\mathcal{U}_{2}%
\mathcal{K}_{\mu \nu }+  \nonumber \\
&&6\mathcal{U}_{1}\mathcal{K}_{\mu \nu }^{2}-6\mathcal{K}_{\mu \nu }^{3})-%
\frac{c_{4}}{2}(\mathcal{U}_{4}g_{\mu \nu }-4\mathcal{U}_{3}\mathcal{K}_{\mu
\nu }+12\mathcal{U}_{2}\mathcal{K}_{\mu \nu }^{2}-24\mathcal{U}_{1}\mathcal{K%
}_{\mu \nu }^{3}+24\mathcal{K}_{\mu \nu }^{4}).  \label{massiveTerm}
\end{eqnarray}%
and the energy-momentum tensor of electromagnetic source in Eq. (\ref{Field
equation}) can be introduced as
\begin{equation}
T_{\mu \nu }=\frac{1}{2}g_{\mu \nu }\mathcal{L}(\mathcal{F})-2\mathcal{L}_{%
\mathcal{F}}F_{\mu \lambda }F_{\nu }^{\lambda }.  \label{Energy momentum}
\end{equation}

\section{Magnetic Solutions in Einstein-Massive Gravity with Maxwell Field}

\label{Sol}

Here, we are going to study the magnetic solutions of Eqs. (\ref{Field
equation}) and (\ref{Maxwell equation}) by considering the Maxwell
electromagnetic field, namely $\mathcal{L}(\mathcal{F})=-\mathcal{F}$. To do
so, we consider the metric of $d-$dimensional spacetime in the following
explicit form
\begin{equation}
ds^{2}=-\frac{\rho ^{2}}{l^{2}}dt^{2}+\frac{d\rho ^{2}}{g(\rho )}%
+l^{2}g(\rho )d\varphi ^{2}+\frac{\rho ^{2}}{l^{2}}%
h_{ij}dx_{i}dx_{j},~~~~~i,j=1,2,3,...,n,  \label{metric}
\end{equation}%
where $g(\rho)$ is an arbitrary function of radial coordinate $\rho $ which
should be determined, $h_{ij}dx_{i}dx_{j}$ is the Euclidean metric on the ($%
d-3$)-dimensional submanifold, and the scale length factor $l$ is related to
the cosmological constant $\Lambda$. In addition, the angular coordinate $%
\varphi$ is dimensionless and ranges in $0\leq \varphi \leq 2\pi $ while $%
x_{i}$'s range is $(-\infty ,+\infty )$. The motivation of considering the
metric gauge [$g_{tt}\varpropto -\rho ^{2}$ and $\left( g_{\rho \rho
}\right) ^{-1}\varpropto g_{\varphi \varphi }$] instead of the usual
Schwarzschild like gauge [$\left( g_{\rho \rho }\right) ^{-1}\varpropto
g_{tt}$ and $g_{\varphi \varphi }\varpropto \rho ^{2}$] comes from the fact
that we are looking for the magnetic solutions instead of electric ones. In
addition, one can obtain such magnetic metric with local transformations $%
t\rightarrow il\varphi $ and $\varphi \rightarrow it/l$ in the horizon flat
Schwarzschild like metric, $ds^{2}=-g(\rho )dt^{2}+\frac{d\rho ^{2}}{g(\rho )%
}+\rho ^{2}d\varphi ^{2}+\frac{\rho ^{2}}{l^{2}}h_{ij}dx_{i}dx_{j}$. In
other words, using such transformation, the metric (\ref{metric}) can be
mapped to $d$-dimensional Schwarzschild like spacetime locally, but not
globally, and therefore, both spacetimes are distinct.

In order to obtain exact solutions, we should make a suitable choice for the
reference metric. Regarding the mentioned local transformation, we consider
the following ansatz for the reference metric
\begin{equation}
f_{\mu \nu }=diag(-\frac{c^{2}}{l^{2}},0,0,\frac{c^{2}}{l^{2}}h_{ij}),
\label{f11}
\end{equation}%
where $c$ in the above equation is a positive constant. Before we go
on, we discuss the reason for considering such a reference metric (\ref{f11}%
). In case of $d$ dimensional black holes, the metric with $(-,+,...,+)$
signature is given by
\begin{equation}
ds^{2}=-g(\rho )dt^{2}+\frac{d\rho ^{2}}{g(\rho )}+\rho ^{2} d\varphi ^{2}+%
\frac{\rho ^{2}}{l^{2}}h_{ij}dx_{i}dx_{j},~~~~~i,j=1,2,3,...,n.
\label{three}
\end{equation}

The black hole solutions in massive gravity are obtained using the
ansatz metric $f_{\mu \nu }=diag(0,0,c^{2},\frac{c^{2}}{l^{2}}h_{ij})$ for
reference metric. In electrical black hole solutions, the metric function, $%
g(\rho )$, is coupled with radial and temporal coordinates whereas in
magnetic spacetime metric (Eq. \ref{metric}), it is coupled with radial and
spatial coordinates. Therefore, to obtain exact solutions in an axially
symmetric spacetime with the form (\ref{metric}), reference metric $f_{\mu
\nu }=diag(0,0,c^{2},\frac{c^{2}}{l^{2}}h_{ij})$ should be modified into $%
f_{\mu \nu }=diag(\frac{-c^{2}}{l^{2}},0,0,\frac{c^{2}}{l^{2}}h_{ij})$. It
should be noted that using the reference metric ($f_{\mu \nu }=diag(\frac{%
-c^{2}}{l^{2}},0,0,\frac{c^{2}}{l^{2}}h_{ij})$), results into a new class of
nontrivial solutions. In addition, it is worth mentioning that this choice
for reference metric, first, cannot produce any infinite value for the bulk
action, since the bulk action contains non-negative powers of $f_{\mu \nu }$%
, and second, it does not preserve general covariance in the transverse
coordinates $t$, $x_{1}$, $x_{2}$, .... .

Using the metric ansatz (\ref{f11}), $\mathcal{U}_{i}$'s can be calculated
in the following forms
\begin{equation}
\mathcal{U}_{1}=\frac{d_{2}c}{\rho },\;\;\mathcal{U}_{2}=\frac{%
d_{2}d_{3}c^{2}}{\rho ^{2}},~~\mathcal{U}_{3}=\frac{d_{2}d_{3}d_{4}c^{3}}{%
\rho ^{3}},~~\mathcal{U}_{4}=\frac{d_{2}d_{3}d_{4}d_{5}c^{4}}{\rho ^{4}},
\label{U}
\end{equation}
where $d_{i}=d-i$. Due to our interest to investigate the magnetic
solutions, we should assume a suitable gauge potential which leads to
consistent field equations
\begin{equation}
A_{\mu }=h(\rho )\delta _{\mu }^{\varphi }.  \label{Gauge potential}
\end{equation}

Using the Maxwell equation (\ref{Maxwell equation}) with $\mathcal{L}(%
\mathcal{F})=-\mathcal{F}$, and the metric (\ref{metric}), one finds the
following differential equation
\begin{equation}
d_{2}F_{\varphi \rho }+\rho F_{\varphi \rho }^{\prime }=0,  \label{Fpr}
\end{equation}%
where $F_{\varphi \rho }=h^{\prime }(\rho )$ and "prime" denotes
differentiation with respect to $\rho $. The solution of Eq. (\ref{Fpr}) is
\begin{equation}
F_{\varphi \rho }=\frac{q}{\rho ^{d_{2}}},  \label{Linfield}
\end{equation}%
where $q$ is an integration constant which may be related to electric
charge. Substituting Eqs. (\ref{metric}) and (\ref{Gauge potential}) in the
field equation (\ref{Field equation}), one can obtain
\begin{eqnarray}
&&\frac{l^{2}}{2}g^{\prime }(\rho )+\frac{\rho }{d_{2}}F_{\varphi \rho }^{2}-%
\frac{d_{3}l^{2}}{2}\left\{ m^{2}\left( \frac{cc_{1}}{d_{3}}+\frac{c_{2}c^{2}%
}{\rho }+\frac{d_{4}c_{3}c^{3}}{\rho ^{2}}+\frac{d_{4}d_{5}c_{4}c^{4}}{\rho
^{3}}\right) -\frac{\left( g(\rho )+\frac{2\Lambda \rho ^{2}}{d_{2}d_{3}}%
\right) }{\rho }\right\} =0, \\
&&  \nonumber \\
&&\frac{l^{2}}{2}g^{\prime \prime }(\rho )+\frac{d_{3}l^{2}}{\rho }g^{\prime
}(\rho )-F_{\varphi \rho }^{2}-\frac{d_{3}l^{2}}{2}\left\{ m^{2}\left( \frac{%
cc_{1}}{\rho }+\frac{d_{4}c_{2}c^{2}}{\rho ^{2}}+\frac{d_{4}d_{5}c_{3}c^{3}}{%
\rho ^{3}}+\frac{d_{4}d_{5}d_{6}c_{4}c^{4}}{\rho ^{4}}\right) -\frac{%
d_{4}g(\rho )}{\rho ^{2}}-\frac{2\Lambda }{d_{3}}\right\} =0.
\end{eqnarray}

Using the above equations, one can calculate the metric function $g(\rho )$
as
\begin{equation}
g(\rho )=\frac{m_{0}}{\rho ^{d_{3}}}-\frac{2\Lambda \rho ^{2}}{d_{1}d_{2}}+%
\frac{2d_{3}q^{2}}{d_{2}\rho ^{2d_{3}}}+m^{2}\left( c_{2}c^{2}+\frac{%
cc_{1}\rho }{d_{2}}+\frac{d_{3}c_{3}c^{3}}{\rho }+\frac{d_{3}d_{4}c_{4}c^{4}%
}{\rho ^{2}}\right) ,  \label{Linmetric}
\end{equation}%
which $m_{0}$ is an integration constant related to the mass. It is
worthwhile to mention that in the absence of massive parameter ($m=0$), the
metric function (\ref{Linmetric}) reduces to the Einstein-standard Maxwell
\cite{OJCDias2}.

\subsubsection{Geometric Properties}

In order to discuss the geometric properties of spacetime, we should focus
on special points of spacetime (such as roots of the metric function) and
boundary of radial coordinate (both $\rho \rightarrow 0$ and $\rho
\rightarrow \infty$) as well.

Since the second term ($\Lambda$ term) of the metric function is dominant
for large values of $\rho$, the asymptotical behavior of the solution (\ref%
{Linmetric}) is adS or dS provided $\Lambda <0$ or $\Lambda >0$.

In order to find the location of curvature singularities (\ref{metric}), one
can calculate the Kretschmann scalar as
\begin{equation}
R_{\mu \nu \lambda \kappa }R^{\mu \nu \lambda \kappa }=\left( \frac{%
d^{2}g(\rho )}{d\rho ^{2}}\right) ^{2}+2d_{2}\left( \frac{1}{\rho }\frac{%
dg(\rho )}{d\rho }\right) ^{2}+2d_{2}d_{3}\left( \frac{g(\rho )}{\rho ^{2}}%
\right) ^{2}.  \label{LinKretschmann}
\end{equation}

Using the metric function (\ref{Linmetric}), it is easy to show that the
Kretschmann scalar (\ref{LinKretschmann}) diverges at $\rho =0$, and
therefore, one may guess that there is a curvature singularity located at $%
\rho =0$, but as we will show, the spacetime will never achieve $\rho =0$.
There are two possible cases for the metric function: first, the metric
function has no root which is interpreted as naked singularity, and second,
the metric function has one or more roots. We assume that $r_{+}$ is the
largest real positive root of the metric function $g(\rho )$. Therefore, the
metric function $g(\rho )$ will be negative for $\rho <r_{+}$ and positive
for $\rho >r_{+}$. This indicates that signature of the metric at
this root changes from $(-,+,+,+,...,+)$ change to $(-,-,-,+,...,+)$. In
general relativity and gravity, although the field equations are metric
dependent, they must not depend on the signature of metric \cite%
{Kossowski,Dray1,Dray2,White}. The mentioned change in the signature of
metric indicates that field equations for $\rho >r_{+}$ and $\rho <r_{+}$
are different resulting into two sets of different metric functions. To avoid
such inconsistency, the possibility of extending the spacetime to $\rho
<r_{+}$ must be removed. To do so, we introduce a new radial coordinate $r$
as
\begin{equation}
r^{2}=\rho ^{2}-r_{+}^{2}\Longrightarrow d\rho ^{2}=\frac{r^{2}}{%
r^{2}+r_{+}^{2}}dr^{2},  \label{coordinate}
\end{equation}%
where $\rho \geq r_{+}$ leads to $r\geq 0$. Applying this coordinate
transformation, the metric (\ref{metric}) should be written as
\begin{equation}
ds^{2}=-\frac{r^{2}+r_{+}^{2}}{l^{2}}dt^{2}+\frac{r^{2}}{\left(
r^{2}+r_{+}^{2}\right) g(r)}dr^{2}+l^{2}g(r)d\varphi ^{2}+\frac{%
r^{2}+r_{+}^{2}}{l^{2}}dX^{2},  \label{change coordinate metric}
\end{equation}%
in which the coordinate $\varphi $\ assumes the value $0\leq \varphi <2\pi $, as usual. The metric function $g(r)$ (Eq. (\ref{Linmetric})) is now given
by
\begin{equation}
g(r)=\frac{m_{0}}{r^{d_{3}}}-\frac{2\Lambda \left( r^{2}+r_{+}^{2}\right) }{%
d_{1}d_{2}}+\frac{2d_{3}q^{2}}{d_{2}r^{2d_{3}}}+m^{2}\left( c_{2}c^{2}+\frac{%
cc_{1}\sqrt{r^{2}+r_{+}^{2}}}{d_{2}}+\frac{d_{3}c_{3}c^{3}}{\sqrt{%
r^{2}+r_{+}^{2}}}+\frac{d_{3}d_{4}c_{4}c^{4}}{\left( r^{2}+r_{+}^{2}\right) }%
\right) ,  \label{change Linmetric}
\end{equation}

The nonzero component of the electromagnetic field in the new coordinates
can be given by
\begin{equation}
F_{\varphi r}=\frac{q}{\left( r^{2}+r_{+}^{2}\right) ^{d_{2}/2}}.
\label{new Linfield}
\end{equation}

One can show that all curvature invariants are functions of $g^{\prime
\prime }$, $g^{\prime }/r$, and\ $g/r^{2}$. Since these terms do not diverge
in the range $0\leq r<\infty $, one finds that all curvature invariants are
finite. Therefore, this spacetime has no curvature singularity and no
horizon. However, the spacetime (\ref{change coordinate metric}) has a conic
geometry and has a conical singularity at $r=0$, because the limit of the
ratio \emph{"circumference/radius"} is not $2\pi$,
\begin{equation}
\lim_{r\longrightarrow 0}\frac{1}{r}\sqrt{\frac{g_{\varphi \varphi }}{g_{rr}}%
}\neq 1.
\end{equation}

The conical singularity can be removed if one exchanges the coordinate $%
\varphi $ with the following period
\begin{equation}
Period_{\varphi }=2\pi \left( \lim_{r\longrightarrow 0}\frac{1}{r}\sqrt{%
\frac{g_{\varphi \varphi }}{g_{rr}}}\right) ^{-1}=2\pi \left( 1-4\mu \right)
,  \label{Period}
\end{equation}%
where $\mu $ is given by%
\begin{equation}
\mu =\frac{1}{4}\left[ 1-\frac{1}{2lr_{0}\left. g^{\prime \prime
}(r)\right\vert _{r=0}}\right] ,  \label{miu}
\end{equation}%
in which $\left. g(r)\right\vert _{r=0}=\left. g^{\prime }(r)\right\vert
_{r=0}=0$, and $\left. g^{\prime \prime }(r)\right\vert _{r=0}$ is
\begin{equation}
\left. g^{\prime \prime }(r)\right\vert _{r=0}=\frac{-2\Lambda }{d_{2}}-%
\frac{2(d_{3})^{2}q^{2}}{d_{2}r_{+}^{d_{1}}}+m^{2}\left( \frac{cc_{1}}{r_{+}}%
+\frac{d_{3}c^{2}c_{2}}{r_{+}^{2}}+\frac{d_{3}d_{4}c_{3}c^{3}}{r_{+}^{3}}+%
\frac{d_{3}d_{4}d_{5}c_{4}c^{4}}{r_{+}^{4}}\right) ,  \label{omega}
\end{equation}%
where shows that the metric (\ref{change coordinate metric}) describes a
spacetime which is locally flat, but has a conical singularity at $r=0$ with
a deficit angle as
\begin{equation}
\delta \phi =8\pi \mu .
\end{equation}

Here, we skip investigation of physical properties of the obtained results.
After obtaining the consequences of nonlinear case, we give a detailed
discussion with comparison.

\section{Magnetic Solutions in the Einstein-Massive Gravity with PMI Field}

In this section, we are going to obtain $d$-dimensional magnetic brane
solutions in the presence of PMI field. Therefore, we consider the PMI
Lagrangian with the following form
\begin{equation}
\mathcal{L}_{PMI}(\mathcal{F})=(-\kappa \mathcal{F})^{s},  \label{PMI}
\end{equation}%
where $\kappa$ and $s$ are coupling and positive arbitrary constants,
respectively. Since the Maxwell invariant is negative in static spacetimes,
hereafter, we set $\kappa =1$ without loss of generality to obtain real
solutions. Also, it is easy to show that when $s$ goes to $1$, the PMI
Lagrangian (\ref{PMI}) reduces to the standard Maxwell Lagrangian ($\mathcal{%
L}_{Maxwell}(\mathcal{F})=- \mathcal{F}$) which we have investigated in the
previous section. It is easy to show that for the case of $power=dimension/4$%
, one can obtain $T^{\mu }_{\mu }=0$ in PMI theory, which is confirmation of
its conformal invariance properties in this case.

Considering Eq. (\ref{PMI}), the electromagnetic field equation (\ref%
{Maxwell equation}) reduces to
\begin{equation}
(2s-1)\rho h^{\prime \prime }(\rho )+d_{2}h^{\prime }(\rho )=0,  \label{Diff}
\end{equation}%
with the following solutions
\begin{equation}
F_{\varphi \rho }=h^{\prime }(\rho )=\left\{
\begin{array}{cc}
\frac{q}{\rho }, & s=d_{1}/2 \\
\frac{(2s-d_{1})q}{(2s-1)\rho ^{d_{2}/(2s-1)}}, & otherwise%
\end{array}%
\right. .  \label{Lagrangian field}
\end{equation}%
where $q$ is an integration constant. Using Eqs. (\ref{metric}) and (\ref%
{Lagrangian field}), one can show that the gravitational field equation (\ref%
{Field equation}) reduces to
\begin{eqnarray}
&&\frac{d_{3}d_{4}g(\rho )}{\rho ^{2}}+\frac{2d_{3}g\prime (\rho )}{\rho }%
+g^{\prime \prime }(\rho )+2\Lambda +\left( 2F_{\varphi \rho }^{2}\right)
^{s}-\frac{d_{3}m^{2}}{\rho ^{4}}\left( c_{1}c\rho ^{3}+d_{4}c_{2}c^{2}\rho
^{2}+d_{4}d_{5}c_{3}c^{3}\rho +d_{4}d_{5}d_{6}c_{4}c^{4}\right) =0, \\
&&  \nonumber \\
&&\frac{d_{2}d_{3}g(\rho )}{\rho ^{2}}+\frac{d_{2}g\prime (\rho )}{\rho }%
+2\Lambda +(1-2s)\left( 2F_{\varphi \rho }^{2}\right) ^{s}-\frac{d_{2}m^{2}}{%
\rho ^{4}}\left( c_{1}c\rho ^{3}+d_{3}c_{2}c^{2}\rho
^{2}+d_{3}d_{4}c_{3}c^{3}\rho +d_{3}d_{4}d_{5}c_{4}c^{4}\right) =0.
\end{eqnarray}

Substituting Eq. (\ref{Lagrangian field}) in the above equations, it is
straightforward to show that the metric function $g(\rho )$ has the
following form
\begin{equation}
g(\rho )=\frac{m_{0}}{\rho ^{d_{3}}}-\frac{2\Lambda \rho ^{2}}{d_{1}d_{2}}%
+m^{2}\left( c_{2}c^{2}+\frac{cc_{1}\rho }{d_{2}}+\frac{d_{3}c_{3}c^{3}}{%
\rho }+\frac{d_{3}d_{4}c_{4}c^{4}}{\rho ^{2}}\right) +\mathbf{A},
\label{metricfun}
\end{equation}%
in which
\[
\mathbf{A}=\left\{
\begin{array}{cc}
\frac{\left( \sqrt{2}q\right) ^{d_{1}}}{\rho ^{d_{3}}}\ln \left( \frac{\rho
}{l}\right) , & s=d_{1}/2 \\
\frac{2^{s}\rho ^{2}(2s-1)^{2}}{d_{2}(2s-d_{1})}\left( \frac{\left(
2s-d_{1}\right) q}{\left( 2s-1\right) \rho ^{d_{2}/(2s-1)}}\right) ^{2s}, &
otherwise%
\end{array}%
\right. .
\]

It is worthwhile to mention that in the absence of massive parameter ($m=0$%
), the metric function (\ref{metricfun}) is just like the metric function
which was obtained before in Ref. \cite{Hendibrane}.

Considering the Kretschmann scalar (\ref{LinKretschmann}), one can show that
the metric (\ref{metric}) with the metric function (\ref{metricfun}), like
the Maxwell case,\ has a singularity at $\rho =0$. However, as we mentioned
before, it is not possible to extend the spacetime to $\rho <r_{+}$ because
of signature changing (see Ref. \cite{MagPMIIII}, for more details). Also,
one can apply the coordinate transformation (\ref{coordinate}) to the metric
(\ref{metric})\ and find the metric function as
\begin{eqnarray}
g(r) &=&\frac{m_{0}}{r^{d_{3}}}-\frac{2\Lambda \left( r^{2}+r_{+}^{2}\right)
}{d_{1}d_{2}}+m^{2}\left( c_{2}c^{2}+\frac{cc_{1}\sqrt{r^{2}+r_{+}^{2}}}{%
d_{2}}+\frac{d_{3}c_{3}c^{3}}{\sqrt{r^{2}+r_{+}^{2}}}+\frac{%
d_{3}d_{4}c_{4}c^{4}}{\left( r^{2}+r_{+}^{2}\right) }\right)  \nonumber \\
&&+\left\{
\begin{array}{cc}
\frac{\left( \sqrt{2}q\right) ^{d_{1}}}{\left( \sqrt{r^{2}+r_{+}^{2}}\right)
^{d_{3}}}\ln \left( \frac{\sqrt{r^{2}+r_{+}^{2}}}{l}\right) , & s=d_{1}/2 \\
\frac{2^{s}\left( r^{2}+r_{+}^{2}\right) (2s-1)^{2}}{d_{2}(2s-d_{1})}\left(
\frac{\left( 2s-d_{1}\right) q}{\left( 2s-1\right) \left( \sqrt{%
r^{2}+r_{+}^{2}}\right) ^{d_{2}/(2s-1)}}\right) ^{2s}, & otherwise%
\end{array}%
\right. ,  \label{change Lagrangian metric}
\end{eqnarray}%
and the electromagnetic field in the new coordinate is
\begin{equation}
F_{\varphi r}=\left\{
\begin{array}{cc}
\frac{q}{\sqrt{r^{2}+r_{+}^{2}}}, & s=(d-1)/2 \\
\frac{(2s-d_{1})q}{(2s-1)\left( \sqrt{r^{2}+r_{+}^{2}}\right) ^{d_{2}/(2s-1)}%
}, & otherwise%
\end{array}%
\right. .
\end{equation}

In this case, like the Maxwell case, this spacetime has a conical
singularity at $r=0$ with the deficit angle $\delta \left( \phi \right)
=8\pi \mu $ where $\mu $ is modified due to the nonlinear electrodynamics
with the following form
\begin{eqnarray}
\left. g^{\prime \prime }(r)\right\vert _{r=0} &=&\frac{-2\Lambda }{d_{2}}%
+m^{2}\left( \frac{cc_{1}}{r_{+}}+\frac{d_{3}c^{2}c_{2}}{r_{+}^{2}}+\frac{%
d_{3}d_{4}c_{3}c^{3}}{r_{+}^{3}}+\frac{d_{3}d_{4}d_{5}c_{4}c^{4}}{r_{+}^{4}}%
\right)   \nonumber \\
&&+\left\{
\begin{array}{cc}
\frac{2^{d_{1}/2}q^{d_{1}}}{r_{+}^{d_{1}}}, & s=d_{1}/2 \\
\frac{2^{s}(2s-1)}{d_{2}}\left( \frac{q^{2}\left( d_{1}-2s\right) ^{2}}{%
(2s-1)^{2}r_{+}^{d_{2}/(2s-1)}}\right) ^{s}, & otherwise%
\end{array}%
\right. .  \label{omegaPMI}
\end{eqnarray}

Due to the complexity of obtained relation in Eq. (\ref{omegaPMI}), it is not
possible to calculate the root and divergence points of deficit angle
analytically, therefore, we study them in some graphs.

\begin{figure}[tbp]
$%
\begin{array}{cc}
\epsfxsize=6.5cm \epsffile{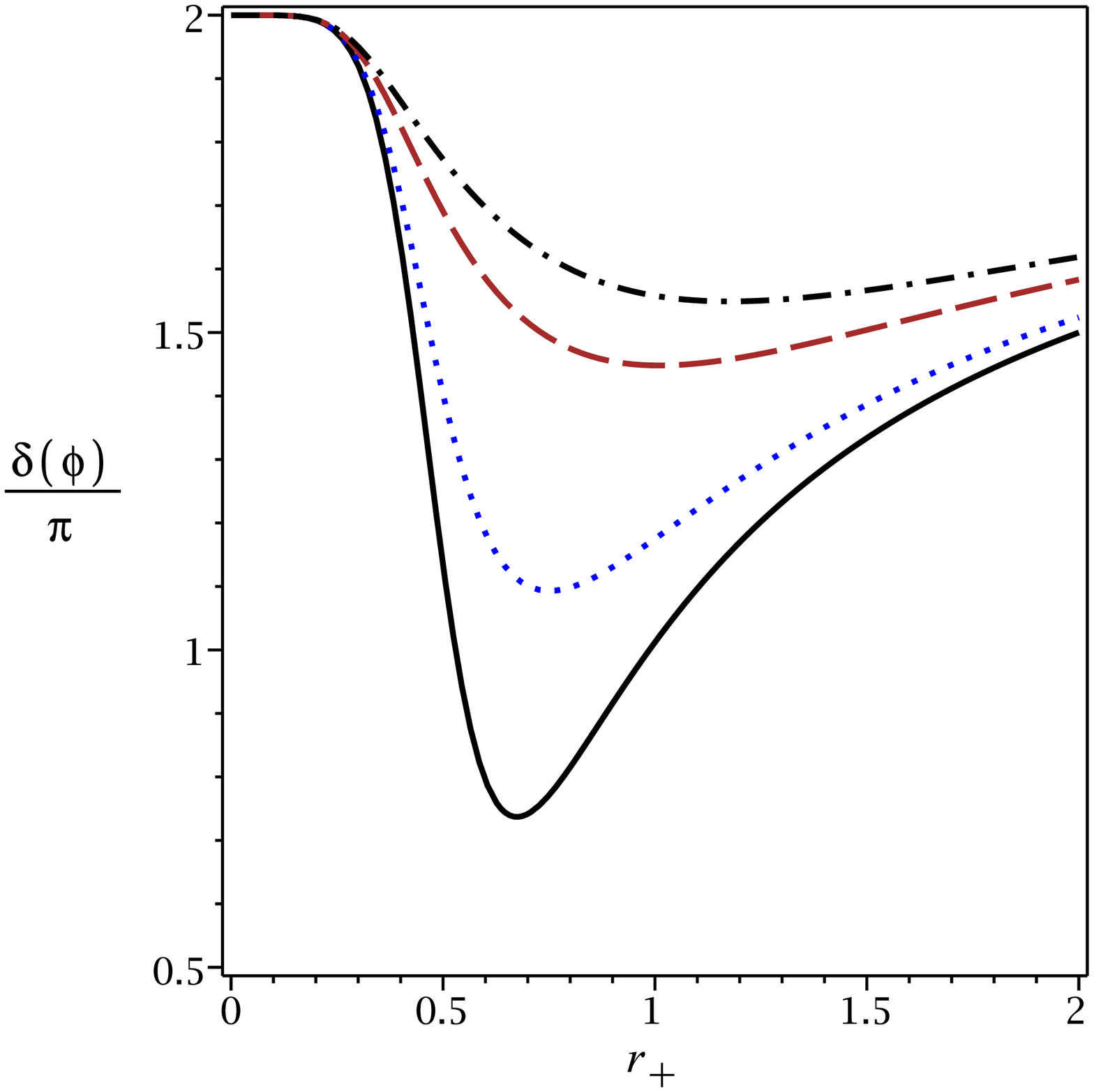} & \epsfxsize=6.5cm %
\epsffile{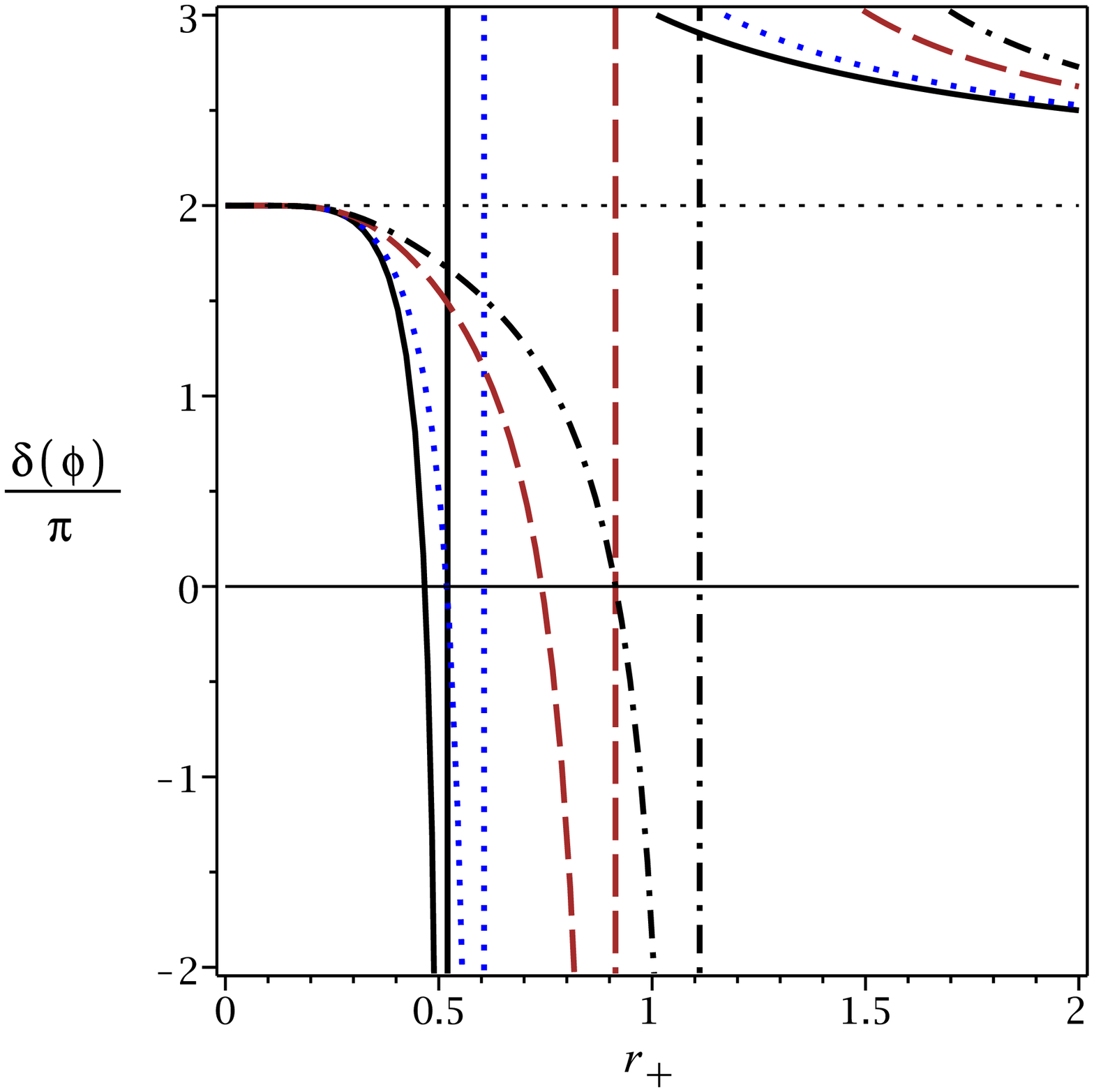}%
\end{array}
$%
\caption{\textbf{\emph{PMI solutions:}} $\protect\delta \left( \protect\phi %
\right) $ versus $r_{+}$ for $q=0.1$, $c=c_{1}=c_{2}=c_{3}=c_{4}=l=1$, $d=5$%
, $s=0.9$, $m=0$ (continuous line), $m=0.4$ (dotted line), $m=0.8$ (dashed
line) and $m=1$ (dashed-dotted line). \newline
\textbf{Left diagram:} $\Lambda =-1$; \textbf{Right diagram:} $\Lambda =1$.}
\label{Fig1}
\end{figure}
\begin{figure}[tbp]
$%
\begin{array}{cc}
\epsfxsize=6.5cm \epsffile{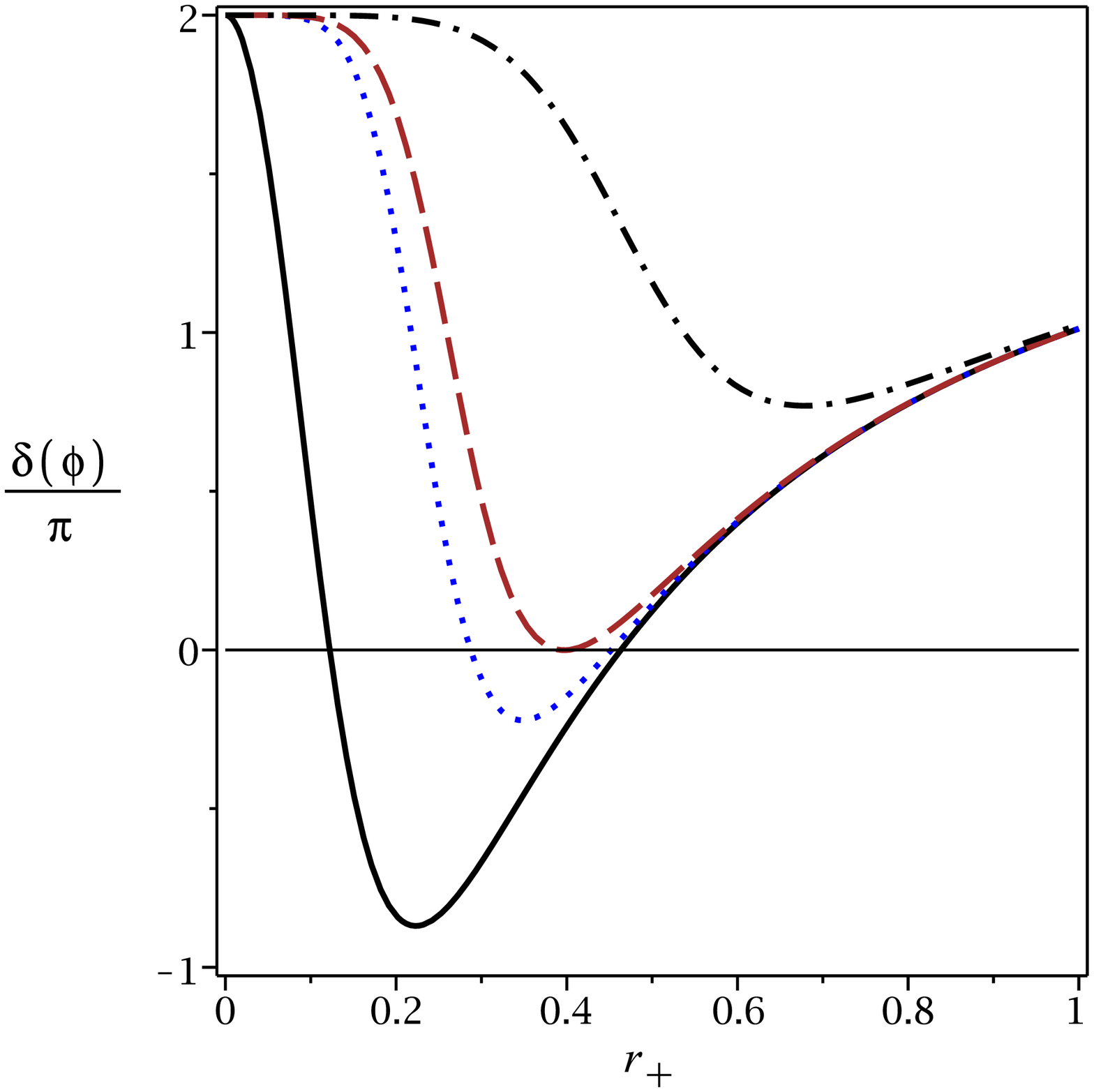} & \epsfxsize=6.5cm %
\epsffile{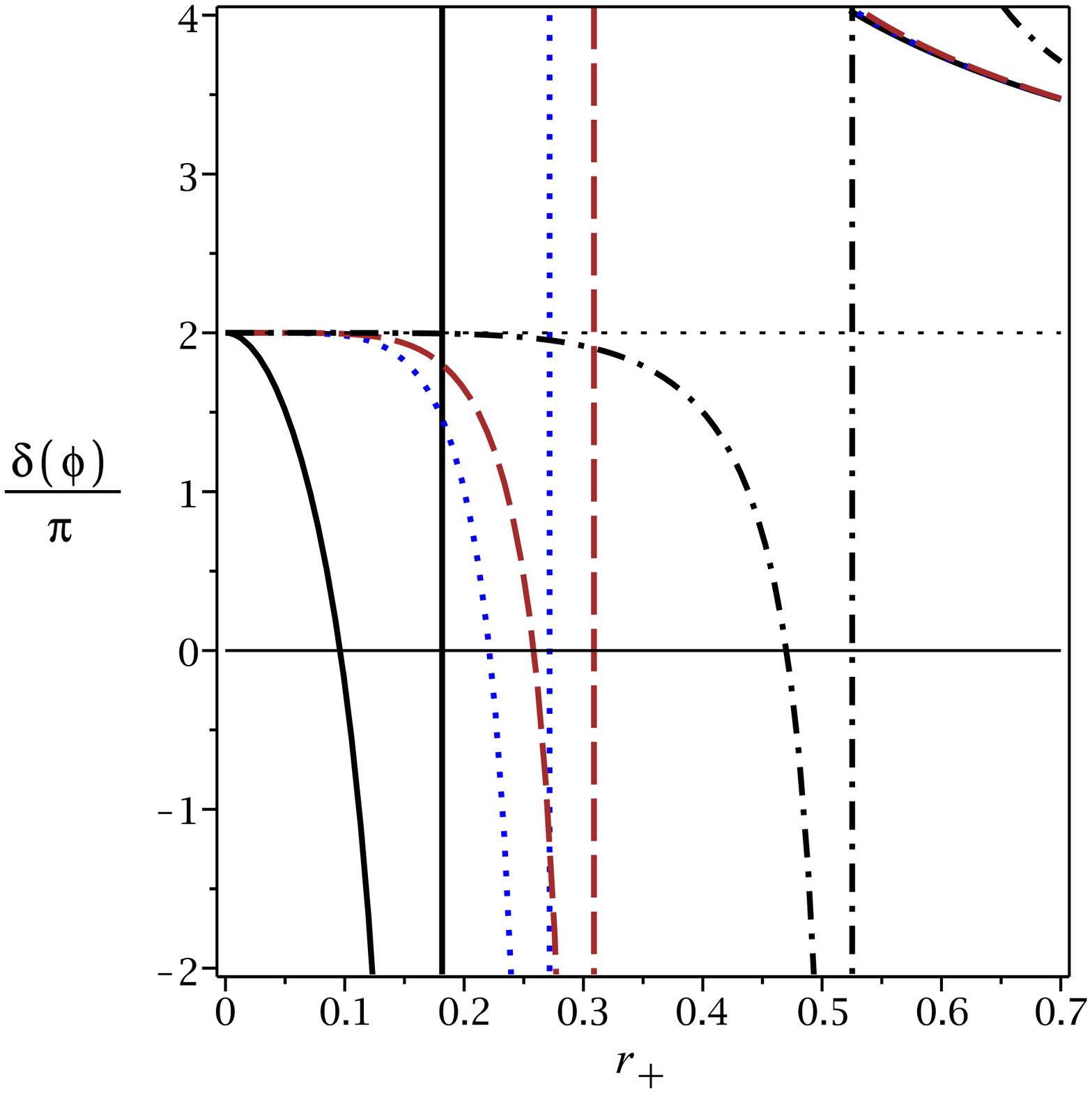}%
\end{array}
$%
\caption{\textbf{\emph{PMI solutions:}} $\protect\delta \left( \protect\phi %
\right) $ versus $r_{+}$ for $c=c_{1}=c_{2}=c_{3}=c_{4}=m=l=1$, $d=5$, $%
s=0.9 $, $q=0$ (continuous line), $q=0.007$ (dotted line), $q=0.0122$
(dashed line) and $q=0.1$ (dashed-dotted line). \newline
\textbf{Left diagram:} $\Lambda =-1$; \textbf{Right diagram:} $\Lambda =1$.}
\label{Fig2}
\end{figure}
\begin{figure}[tbp]
$%
\begin{array}{cc}
\epsfxsize=6.5cm \epsffile{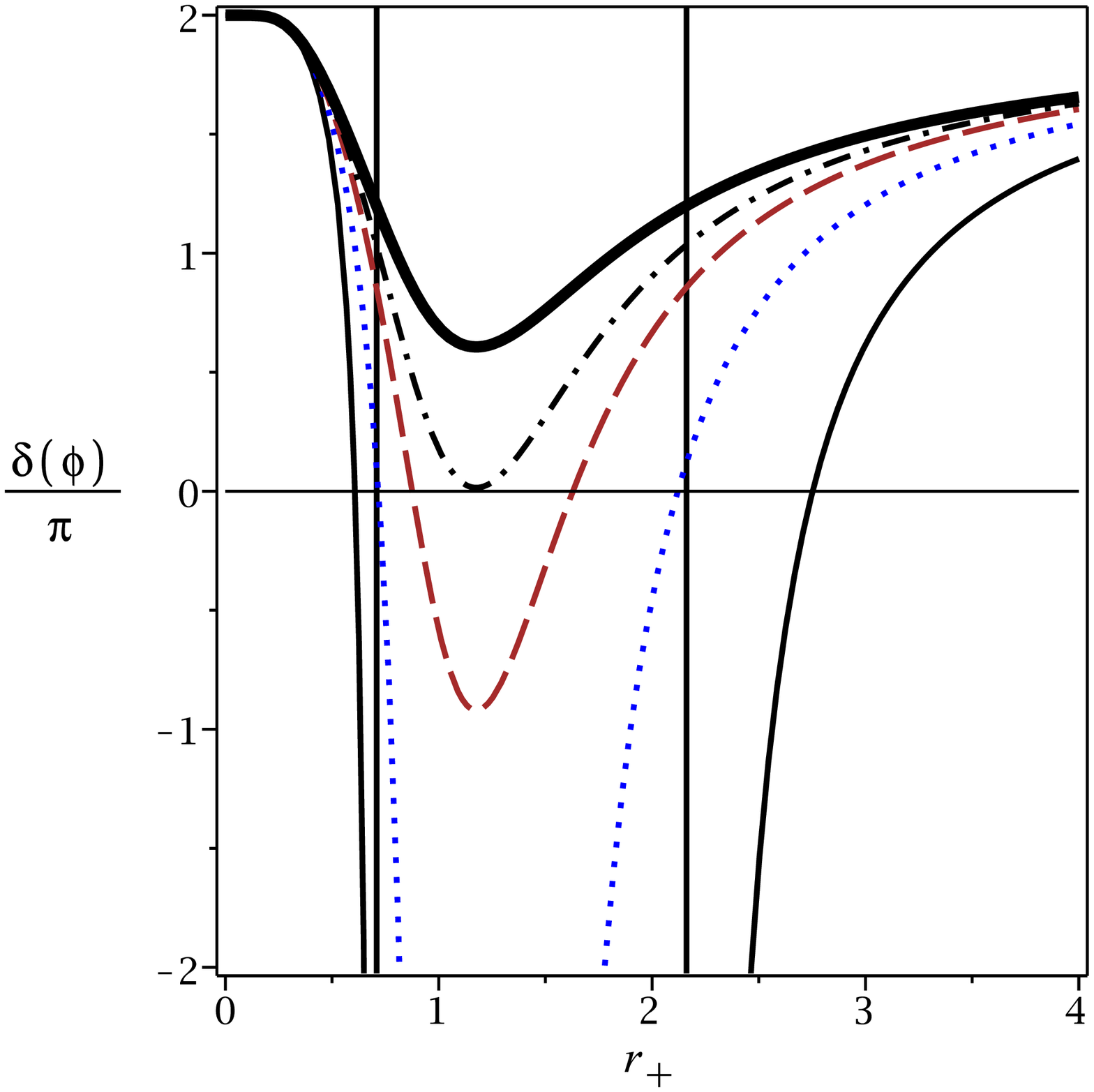} & \epsfxsize=6.5cm %
\epsffile{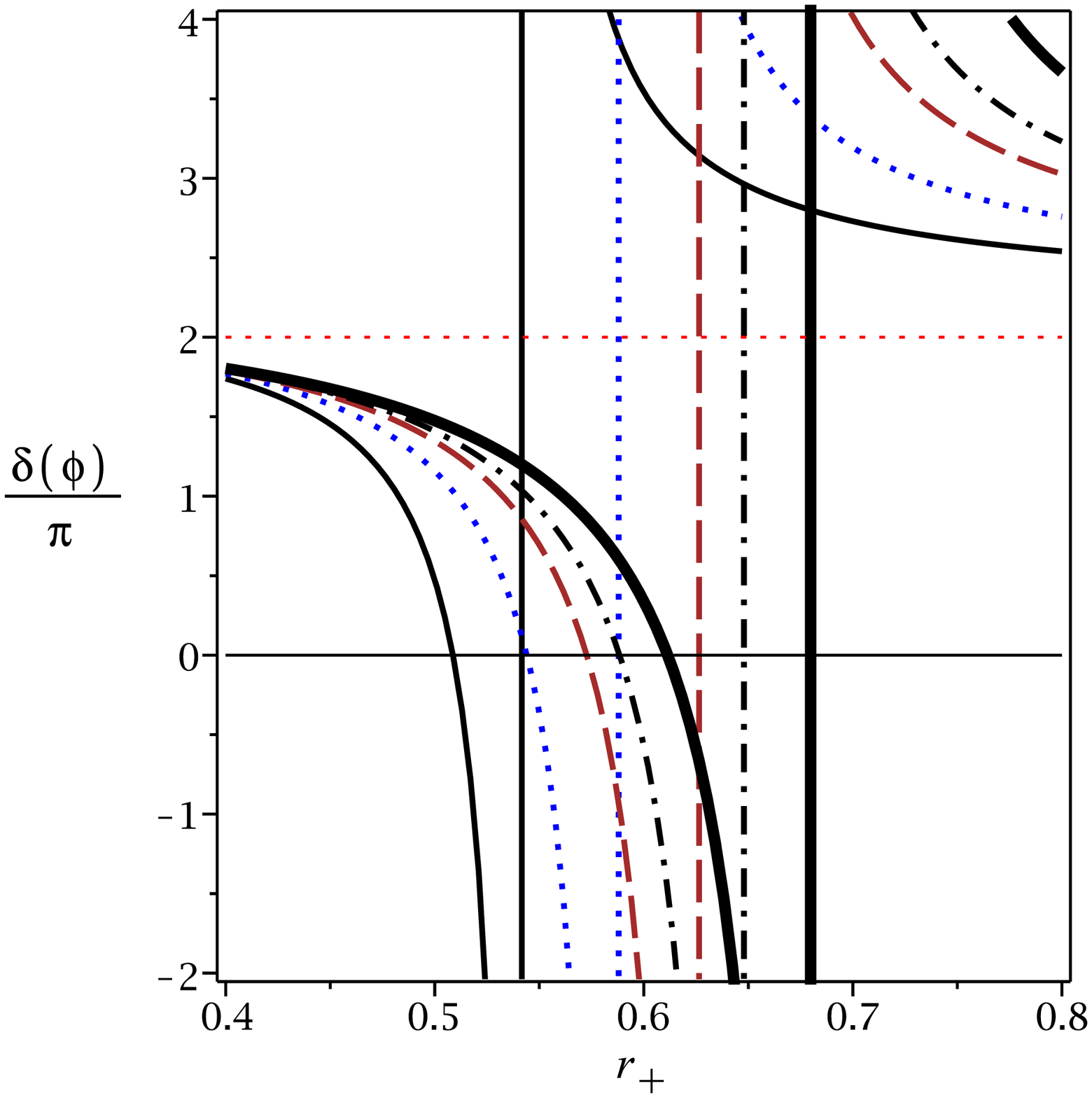}%
\end{array}
$%
\caption{\textbf{\emph{PMI solutions:}} $\protect\delta \left( \protect\phi %
\right) $ versus $r_{+}$ for $q=0.1$, $c=c_{2}=c_{3}=c_{4}=m=l=1$, $d=5$, $%
s=0.9$, $c_{1}=-10$ (continuous line), $c_{1}=-7.87$ (dotted line), $%
c_{1}=-6.5$ (dashed line), $c_{1}=-5.855$ (dashed-dotted line) and $c_{1}=-5$
(bold line). \newline
\textbf{Left diagram:} $\Lambda =-1$; \textbf{Right diagram:} $\Lambda =1$.}
\label{Fig3}
\end{figure}
\begin{figure}[tbp]
$%
\begin{array}{cc}
\epsfxsize=6.5cm \epsffile{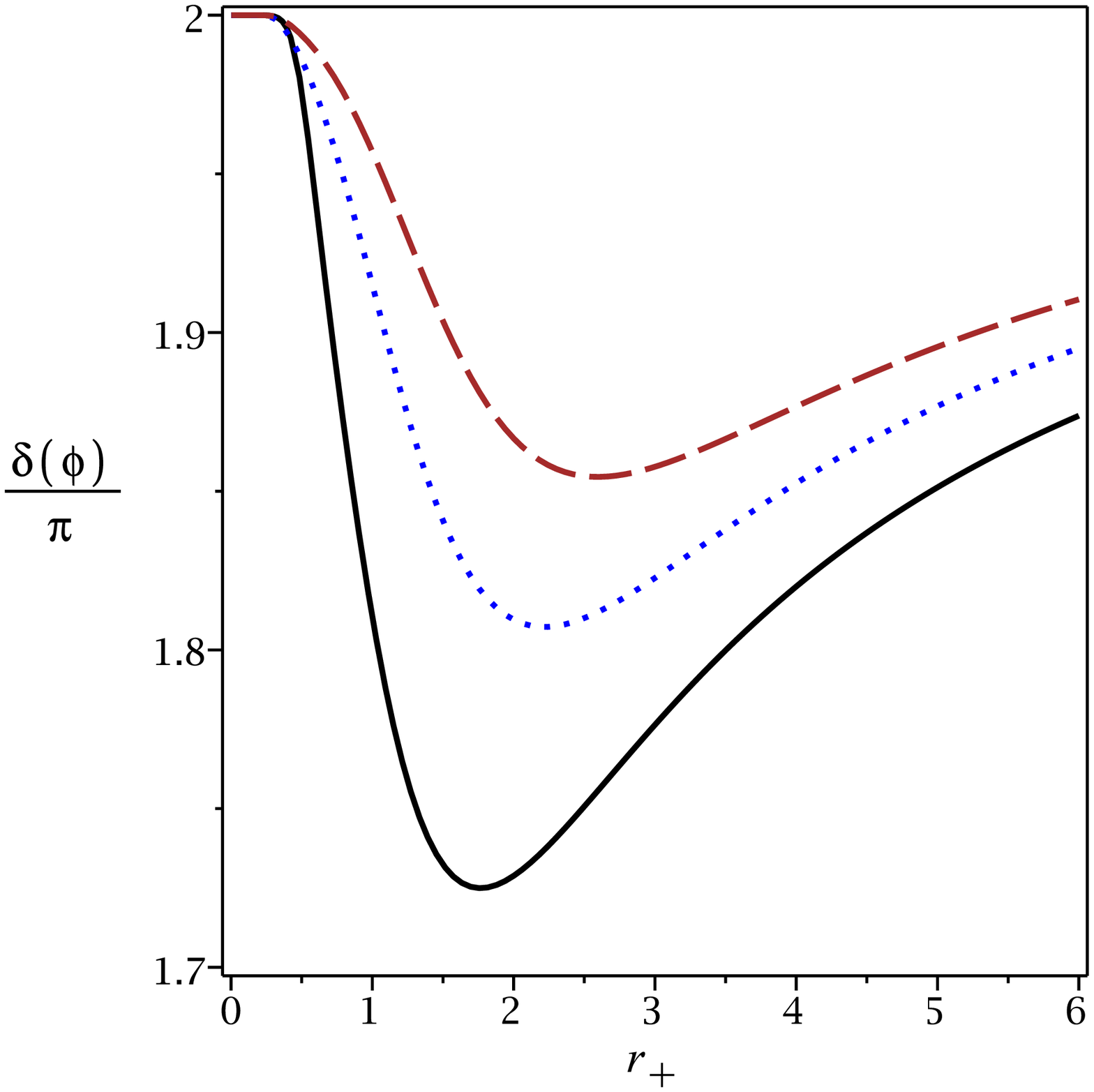} & \epsfxsize=6.5cm %
\epsffile{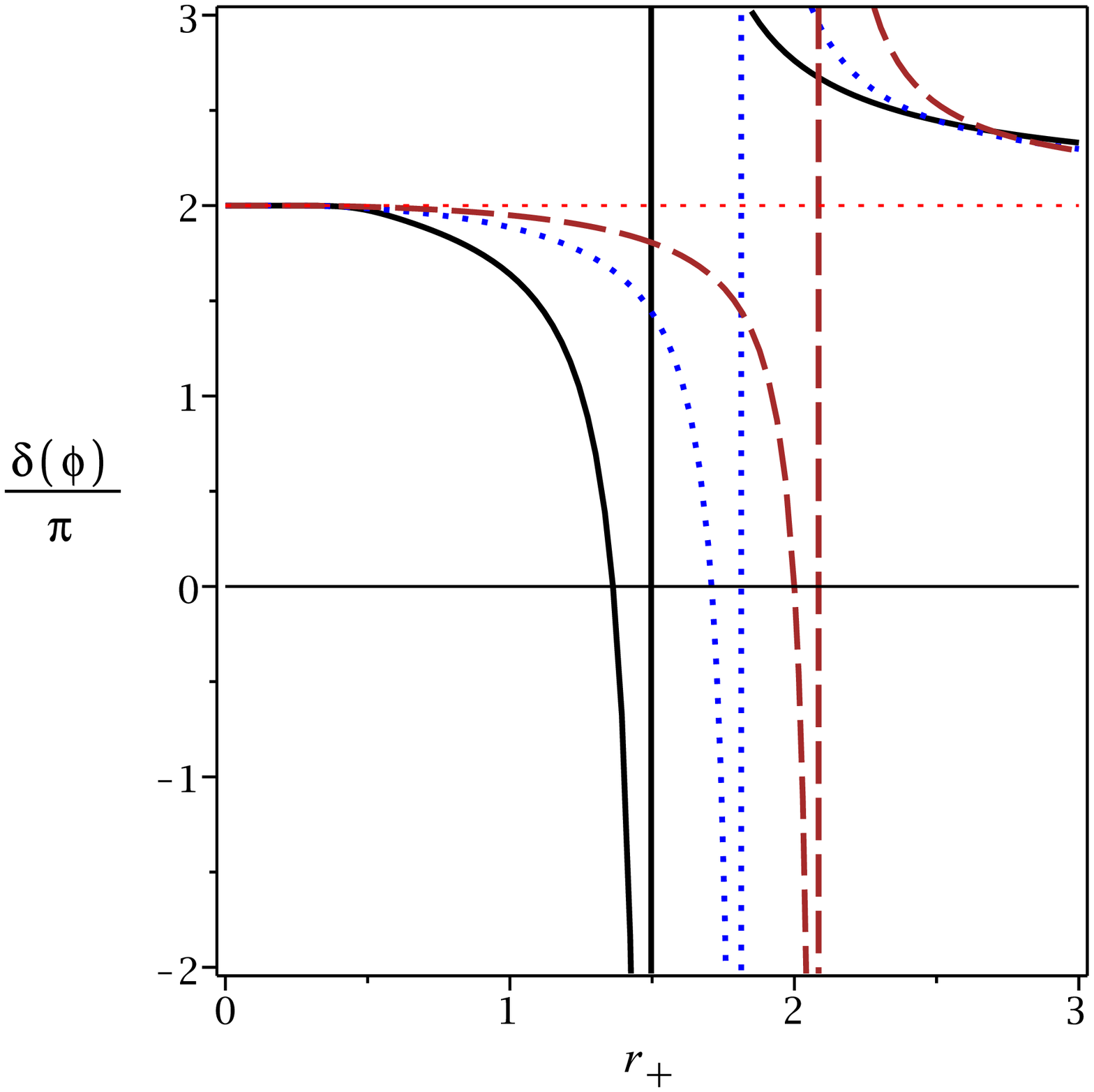}%
\end{array}
$%
\caption{\textbf{\emph{PMI solutions:}} $\protect\delta \left( \protect\phi %
\right) $ versus $r_{+}$ for $q=0.1$, $c=c_{1}=c_{2}=c_{3}=c_{4}=m=l=1$, $%
d=5 $, $s=0.8$ (continuous line), $s=1$ (dotted line) and $s=1.1$ (dashed
line). \newline
\textbf{Left diagram:} $\Lambda =-1$; \textbf{Right diagram:} $\Lambda =1$.}
\label{Fig4}
\end{figure}
\begin{figure}[tbp]
$%
\begin{array}{cc}
\epsfxsize=6.5cm \epsffile{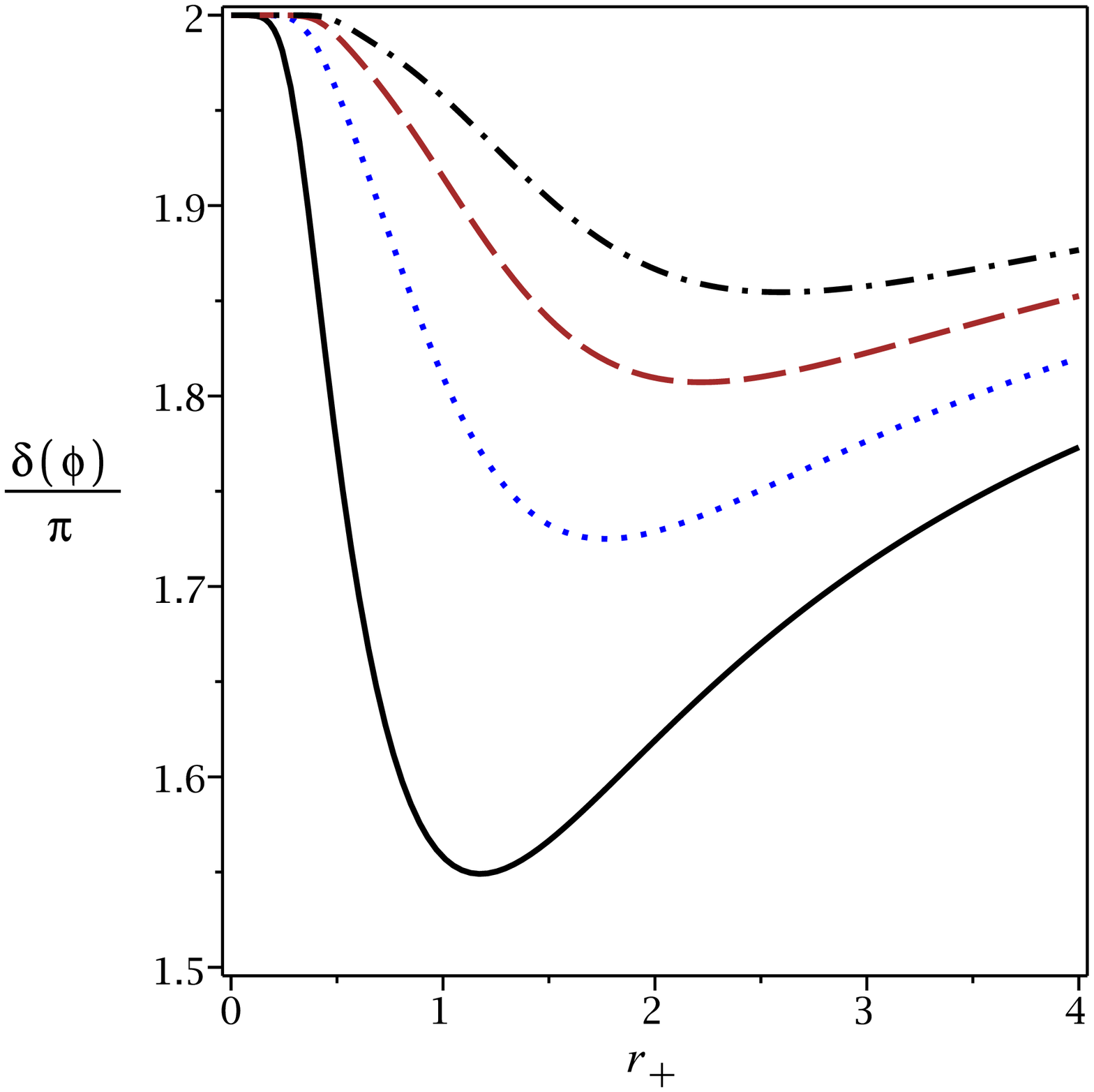} & \epsfxsize=6.5cm %
\epsffile{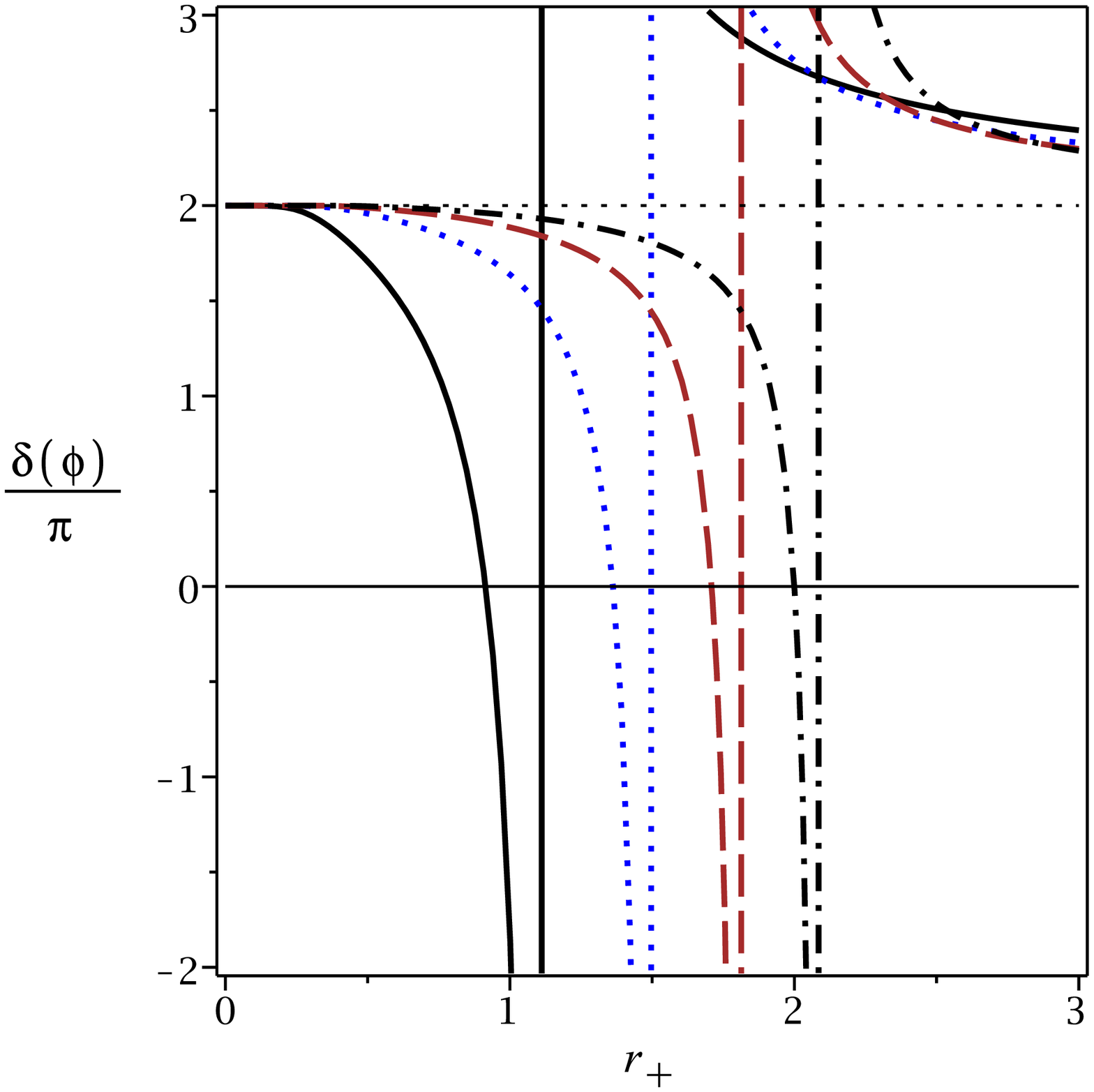}%
\end{array}
$%
\caption{\textbf{\emph{PMI solutions:}} $\protect\delta \left( \protect\phi %
\right) $ versus $r_{+}$ for $q=0.1$, $c=c_{1}=c_{2}=c_{3}=c_{4}=m=l=1$, $%
s=0.9$, $d=5$ (continuous line), $d=6$ (dotted line), $d=7$ (dashed line)
and $d=8$ (dashed-dotted line). \newline
\textbf{Left diagram:} $\Lambda =-1$; \textbf{Right diagram:} $\Lambda =1$.}
\label{Fig5}
\end{figure}

Before starting, we should point it out that we have an upper limit of $%
-\infty <\delta \phi \leq 2\pi$ on the values that deficit angle can
acquire. This limit is marked with a horizontal dotted line in plotted
diagrams. The value of deficit angle determines the geometrical structure of
solutions. Depending on geometrical properties, gravitational effects and
lensing properties of the magnetic solutions, hence topological defects will
be different. Here, we see that depending on choices of different
parameters, deficit angle could be positive/negative and it may have roots
and divergence points. In order to highlight the effects of background
spacetime, we have plotted two series of diagrams for AdS (left panels of
Figs. \ref{Fig1}-\ref{Fig5}) and dS (right panels of Figs. \ref{Fig1}-\ref%
{Fig5}).

Evidently, for AdS case, depending on the choices of different parameters,
deficit angle could have: I) two roots in which between roots, the deficit
angle negative valued whereas before smaller and after larger roots, it is
positive. II) one extreme root in which the deficit angle is always positive
valued. III) two roots with one divergency where between smaller/larger root
and divergency the deficit angle is negative and everywhere else, it is
positive valued. IV) finally, two roots with two divergencies in which the
divergencies are located between the roots. In this case, between smaller
(larger) root and smaller (larger) divergency, the deficit angle is negative
valued. Between divergencies, it is positive but its values are not in
permitted area. Only before (after) smaller (larger) root, the deficit angle
is positive valued and within permitted area.

On contrary, for dS case, plotted diagrams show existence of a root and a
divergency for deficit angle. Before root and after divergency, the deficit
angle is positive where only before root, permitted values of the deficit
angle exists whereas after divergency its values are not within permitted
ones.

The number of roots is a decreasing function of the mass of graviton ($m$)
(left panel of Fig. \ref{Fig1}), electric charge (left panel of Fig. \ref%
{Fig2}), $c_{1}$ (left panel of Fig. \ref{Fig3}), nonlinearity parameter
(left panel of Fig. \ref{Fig4}) and dimensions (left panel of Fig. \ref{Fig5}%
) for AdS case. On the other hand, for dS case, the places of root and
divergency are increasing functions of $m$ (right panel of Fig. \ref{Fig1}),
$q$ (right panel of Fig. \ref{Fig2}), $c_{1}$ (right panel of Fig. \ref{Fig3}%
), $s$ (right panel of Fig. \ref{Fig4}) and dimensions (right panel of Fig. %
\ref{Fig5}).

The existence of positive valued deficit angle results into conic like
geometrical structure for our astrophysical objects, hence topological
defects are known as horizonless magnetic solutions. On contrary, the existence
of negative values of deficit angle leads to a saddle-like cone structure
for the solutions. These two different structures for magnetic solutions
could be related to different second fundamental form of spacetime. On the
other hand, it was argued that positivty/negativity of the deficit angle
results into attractive-type/repulsive-type gravitational potentials
(furthers details could be found in Refs. \cite{neg1,neg2,neg4,neg5}).

Considering different geometrical structure depending on the sign of deficit
angle, one can conclude that the root of deficit angle is where magnetic
solutions have phase transition-like behavior. In other words, since there
is a change of sign at the root of deficit angle, magnetic solutions go
under a typical topological phase transition in these points. It could be
pointed out that there are cases in which roots are extreme ones. In these
cases, although no change of sign takes place, the total geometrical
structure of the solutions presents diverse different comparing to the
non-zero deficit angle (absence of conic like singularity for zero deficit
angle). Therefore, it could be stated that extreme roots are also marking
phase transition points. Another point which carries the properties of phase
transition for magnetic solutions is divergency of the deficit angle. In
other words, divergencies of the deficit angle could be interpreted as
places in which magnetic solutions go under a phase transition. This is due
to the fact that deficit angle has smooth behavior everywhere except at
divergencies which are discontinuities. Usually, around these divergence
points, the sign of deficit angle is changed. In other words, there is a
change in the sign of deficit angle before and after divergence point.

Although different parameters have specific contributions in
existence/absence of root and divergency for deficit angle, the highest
effects belong to the $\Lambda$ term, hence structure of the background
spacetime.

For dS spacetime (positive $\Lambda$), existence of both points (root and
divergence) irrespective of different parameters is evident. Before the
divergency, the values of deficit angle are within permitted area while
after it, the values are in forbidden region. The root of deficit angle in
this case is located at the permitted area. Therefore, one can state that
for dS case, the existence of deficit angle is limited to region before its
divergency and in this region, deficit angle enjoys a phase transition
related to the existence of root. The length of permitted region for deficit
angle is a function of massive parameters, electric charge, dimensions and
nonlinearity parameter.

For AdS spacetime, the situation is different. Existence of divergency
depends on positivity and negativity of massive coefficient $c_{1}$ and it
is found for sufficiently small and negative values of this parameter.
Interestingly, contrary to dS case, AdS spacetime could enjoy the existence
of up to two divergencies in its deficit angle (for sufficiently small and
negative $c_{1}$). In the case of one divergence point, the divergency
exists between two roots and signature of the deficit angle around it is the
same (it is negative). In this case, the deficit angle enjoys two roots and
one divergency. For the case of two divergencies, the divergence points are
between two roots. Around divergencies the sign of deficit angle changes.
Between the divergencies, the deficit angle is positive valued but within
prohibited region. Therefore, the magnetic solutions have phase transition
over a region which is marked with divergencies. This shows that in this
case, the deficit angle has two roots and one divergency with a prohibited
region. The study here showed that generalization to massive gravity
introduces some new phase transitions into magnetic solutions. This
highlights the effects of the massive gravity in geometrical structure of
the solutions, hence their physical properties.

\section{Conclusions}

The paper was dedicated to study the nonlinearly charged magnetic brane
solutions in the presence of massive gravity. The exact solutions were
obtained and the absence of black hole solutions was confirmed. The
existence of conic like singularity was shown and it was pointed out that
geometrical, hence, physical/gravitational properties of the solutions
depend on a value known as deficit angle.

This property of the solutions (deficit angle) determines the total
structure of magnetic branes. There is a diverse difference in the
geometrical properties of the solutions with positive deficit angle
comparing to negative ones. These geometrical properties are providing
guidelines for how phenomena such as lensing property would be different.
That being said, roots and divergencies could be interpreted as topological
phase transition points. In roots, the transition is being done smoothly
while in the divergencies, system jumps between different deficit angles,
hence geometrical structure.

In general, it was shown that existence of the divergencies for
deficit angle were the background spacetime and massive gravity dependent.
If the massive coefficients are positive valued, only for dS background,
deficit angle could acquire divergency whereas, the AdS case enjoys only
root in its deficit angle. On the contrary, if the massive coefficients
could be negative, for both AdS and dS backgrounds, it is possible to
introduce multi geometrical phase transition and a prohibited region.
Existence of the prohibited regions indicates that our magnetic solutions
are bounded by specific limits. These limiting areas and the conditions for
them are rooted in massive gravity and its coefficients. Despite the effects
of other parameters on limiting areas and the conditions, in the absence of
massive coefficients, these limiting areas would rather vanish or
significantly be modified. The effects of nonlinearity nature of the
solutions in the case of AdS spacetime was in level of modifying the number
of roots. Whereas for dS spacetime, it was only in level of modifying the
prohibited/permitted region for deficit angle.

The obtained solutions here contain magnetic brane ones. Considering the AdS
nature of the solutions and their phase transitions, it is possible to
conduct studies in the context of AdS/CFT correspondence. Furthermore, one
can investigate trajectory of the particles and lensing properties of these
solutions in more details to understand the effects of massive gravity and
nonlinear electromagnetic fields. In addition, it is notable that our
solutions are static and independent of time. One may modify these solutions
to the case of dynamic time dependent for investigating the
"self-acceleration" properties \cite{self1,self2}. We leave these subjects
for the future works.

\begin{acknowledgements}
We would like to thank the referee for his/her insightful
comments. The authors wish to thank Shiraz University Research
Council. This work has been supported financially by Research
Institute for Astronomy and Astrophysics of Maragha.
\end{acknowledgements}

\end{document}